# Demonstration of a two-dimensional $\mathcal{PT}$-symmetric crystal: Bulk dynamics, topology, and edge states

Mark Kremer[1], Tobias Biesenthal[1], Matthias Heinrich[1], Ronny Thomale[2], and Alexander Szameit[1]

[1]Institute for Physics, University of Rostock, Albert-Einstein-Straße 23, 18059 Rostock, Germany

[2]Department of Physics and Astronomy, Julius-Maximilians-Universität Würzburg, Am Hubland, 97074 Würzburg, Germany

**In 1998, Carl Bender challenged the perceived wisdom of quantum mechanics that the Hamiltonian operator describing any quantum mechanical system has to be Hermitian [1]. He showed that Hamiltonians that are invariant under combined parity-time ($\mathcal{PT}$) symmetry transformations likewise can exhibit real eigenvalue spectra [2,3]. These findings had a particularly profound impact in the field of photonics, where $\mathcal{PT}$-symmetric potential landscapes can be implemented by appropriately distributing gain and loss for electromagnetic waves [4,5]. Following this approach, it was possible to show some of the hallmark features of $\mathcal{PT}$ symmetry, such as the existence of non-orthogonal eigenmodes [6], non-reciprocal light evolution [7], diffusive coherent transport [8], and to study their implications in settings including $\mathcal{PT}$-symmetric lasers [9,10] and topological phase transitions [11,12]. Similarly, $\mathcal{PT}$-symmetry has enriched other research fields ranging from $\mathcal{PT}$-symmetric atomic diffusion [13], superconducting wires [14,15], and $\mathcal{PT}$-symmetric electronic circuits [16]. Nevertheless, to this date, all experimental implementations of $\mathcal{PT}$-symmetric systems have been restricted to one dimension, which is mostly due to limitations in the technologies at hand for realizing appropriate non-Hermitian potential landscapes. In this work, we report on the experimental realization and characterization of a two-dimensional $\mathcal{PT}$-symmetric system by means of photonic waveguide lattices with judiciously designed refractive index landscape with alternating loss. A key result of our work is the demonstration of a non-Hermitian two-dimensional topological phase transition that coincides with the emergence of mid-gap edge states. Our findings pave the grounds for future**

**investigations exploring the full potential of 𝒫𝒯-symmetric photonics in higher dimensions. Moreover, our approach may even hold the key for realizing two-dimensional 𝒫𝒯-symmetry also in other systems beyond photonics, such as matter waves and electronics.**

𝒫𝒯-symmetric systems are described by a Hamiltonian that is invariant under parity-time symmetry transformations [1]. In a more formal language, this means that if the Hamiltionian $\hat{H}$ commutes with the 𝒫𝒯-operator: $[\hat{H}, PT] = 0$, and the Hamiltonian shares the same set of eigenstates with the 𝒫𝒯-operator, then the entire set of of $\hat{H}$ is real. A necessary condition for this symmetry to hold is that the underlying potential obeys the relation $\hat{V}(-x) = \hat{V}^*(x)$ [1]. Whereas complex potentials tend to be difficult to realise in most physical systems, in 2007 it was shown that photonics provides a suitable testbed due to the complex-valued character of the refractive index [4,5]. Since then, 𝒫𝒯-symmetric systems have been explored in a variety of photonics platforms, ranging from waveguide arrays [6], fiber lattices [7] and coupled optical resonators [17] to plasmonics [18] and microwave cavities [19]. The implementation of 𝒫𝒯-symmetry in photonics is based on the observation that the Schrödinger equation of quantum mechanics for the probability density $\psi(x, y, t)$

$$i\hbar \frac{\partial}{\partial z}\psi(x,y,t) = -\frac{\hbar^2}{2m}\nabla^2 \psi(x,y,t) + V(x,y,t)\psi(x,y,t) \tag{1}$$

and the paraxial Helmholtz equation of electromagnetism for the electric field amplitude $E(x, y, z)$

$$i\frac{n_0}{2k_0^2}\frac{\partial}{\partial z}E(x,y,z) = -\frac{n_0}{2k_0^2}\nabla^2 E(x,y,z) - n(x,y,z)E(x,y,z) \tag{2}$$

are formally equivalent if the potential $\hat{V}(x)$ in the Schrödinger equation is replaced by the refractive index profile $-n(x)$ in the Helmholtz equation [20]. 𝒫𝒯-symmetry then translates into the condition for the complex refractive index

$$n(-x) = n^*(x). \tag{3}$$

In other words, the real part $\Re\{n(x)\}$ needs to follow a symmetric distribution, while the imaginary part $\Im\{n(x)\}$ has to be antisymmetric under the parity operation. In general, the latter implies that loss in one propagation direction has to be substituted by an identical gain in the opposite direction [4]. It turns out, however, that this stringent requirement can be relaxed in tight-binding systems, where an alternating loss distribution is sufficient to obtain $\mathcal{PT}$-symmetric behavior [21,8]. Indeed, such passive systems exhibit exactly the same evolution dynamics that one would expect in in active structures if one accounts for a constant global loss by normalizing the field intensity.

In our work, we consider so-called "photonic graphene", a regular arrangement of waveguides in a honeycomb geometry (sketched in Fig. 1a) [22]. In order to implement the necessary potential condition for $\mathcal{PT}$–symmetry Eq. (3), the two triangular sublattices of the graphene may exhibit different loss, symbolized by the yellow and blue filling of the individual lattice sites, respectively. Launching a light beam into the waveguides results in spatial beam dynamics (governed by Eq. (2)) that, for the case of identical loss in both sublattices, resembles the evolution of a single electron in carbon-based graphene according to Eq. (1). One of the striking features of the graphene band structure is the existence of the so-called Dirac region in the vicinity of the conical intersection points ("diabolical points") between the first and the second bands, displayed in Fig. 1b. In these regions, the tight-binding Hamiltonian of our $\mathcal{PT}$–symmetric photonic graphene can be expanded into a Taylor series [23] to obtain a mathematical structure resembling the relativistic Dirac equation, which describes relativistic quantum particles:

$$\hat{H} = c\sqrt{3}\tilde{v}\sigma_1 + c\sqrt{3}\tilde{\mu}\sigma_2 + i\gamma\sigma_3 \qquad (4)$$

Here, $\sigma_{1,2,3}$ are the Pauli matrices:

$$\sigma_1 = \begin{pmatrix} 0 & 1 \\ 1 & 0 \end{pmatrix}; \; \sigma_2 = \begin{pmatrix} 0 & -i \\ i & 0 \end{pmatrix}; \; \sigma_3 = \begin{pmatrix} 1 & 0 \\ 0 & -1 \end{pmatrix}; \qquad (5)$$

$2\gamma$ denotes the difference in the loss between the sublattices, and $c$ is the intersite hopping. The quantities $\tilde{\mu}$ and $\tilde{v}$ represent the components of the transverse wave vector $k_x, k_y$ measured from the position of the original Dirac point. For simplicity, we suppressed an additional term $-i\Gamma\sigma_0$ that arises from the passive nature of our system, where $\Gamma$ is the average loss of both sublattices, and $\sigma_0$ is the unity matrix. The non-Hermitian Hamiltonian in Eq. (4) exhibits a complex dispersion relation with a non-real eigenvalue spectrum [24]. Mathematically, complex eigenvalues of the Hamiltonian appear whenever the $\mathcal{PT}$-operator and the Hamiltonian cease to share all of their eigenvectors. Such a system is said to have broken $\mathcal{PT}$-symmetry, although the $\mathcal{PT}$-operator still commutes with the Hamiltonian. This seeming paradox stems from the fact that the $\mathcal{T}$-operator is anti-linear. A graph of the real part of the graphene dispersion relation with $\frac{\gamma}{c} = 0.32$ is shown in Fig. 1c. In contrast to "conventional" (i.e., dissipationless) graphene, the real part of the dispersion relation is now a single-sheeted hyperboloid. The lower part of Fig. 1c shows the imaginary part of the dispersion relation, highlighting the purely imaginary eigenvalues around the original vertices.

One can drive the system back into the unbroken $\mathcal{PT}$–symmetry regime by applying a linear strain $\tau$. This strain is applied as indicated in Fig. 1a by the red connections between the atoms, where $\tau = 1$ corresponds to the unstrained case. In Hermitian lattices ($\gamma = 0$), increasing the strain pushes pairs of Dirac points towards one another until they merge at $\tau = 2$ [25]. For any given loss factor $\gamma$, all eigenvalues of the system become real above a threshold strain $\tau \geq 2 + \frac{\gamma}{c}$ [23]. Therefore, in such a setting, the structure exhibits unbroken $\mathcal{PT}$–symmetry, with the transition occurring exactly at $\tau = 2 + \frac{\gamma}{c}$. The Hamiltonian of this system reads as [3]

$$\widehat{H} = \left[c(\tau - 2) - \frac{3}{2}ct\tilde{\mu}^2 + c\tilde{v}^2\right]\sigma_1 + ct\sqrt{3}\tilde{\mu}\sigma_2 + i\gamma\sigma_3, \quad (6)$$

with $\Delta = c(t - 2)$ denoting the gap in the spectrum [23]. In Fig. 1d, we show the real and imaginary parts of the dispersion relation for photonic graphene with the same loss factor as in Fig. 1c in the presence of a strain given by $\tau = 2 + \frac{\gamma}{c} + 0.61$. Evidently, all eigenvalues are real, and a gap has opened in agreement with Eq. (6).

In order to implement the $\mathcal{PT}$-symmetric photonic graphene lattice, we employ the direct laser-writing technology [26]. The desired loss in the system is realized by introducing a certain concentration of microscopic scattering points along the waveguides by dwelling, as sketched in Fig. 2a. As the dwelling time and the separation between the scattering points can be individually tuned, this approach allows for a wide range of artificial losses to be chosen without compromising the real part of the refractive index or introducing directionality into the system. Figure 2b shows our calibration measurements of the realized loss, as a function of dwelling time and scattering point separation. The general trends are clearly visible: The smaller the separation and the longer the dwelling times, the higher is the cumulative loss of the light propagating through the waveguides.

With these tools at hand, we can now proceed to experimentally demonstrate the $\mathcal{PT}$-symmetry transition in our two-dimensional photonic graphene lattice. When no strain is present, $\mathcal{PT}$-symmetry is broken and, hence, the spectrum is complex. Moreover, the light resides predominantly in the waveguides without loss [27]. As the strain is increased, however, the system moves into the unbroken $\mathcal{PT}$-symmetry regime, resulting in a real spectrum. In the broken $\mathcal{PT}$-phase, the eigenvalues exhibit different imaginary parts, and the power remaining in the lattice depends strongly on which waveguide the light is injected to. In contrast, in the unbroken $\mathcal{PT}$-phase all eigenvalues exhibit the same imaginary part and, for an infinte system, the power decay in the lattice were independent of the excited waveguide. Therefore, as the strain is increased above the critical value of $\tau = 2 + \frac{\gamma}{c}$, the standard deviation of the transmitted power will eventually vanish [12]. We demonstrate this behavior by fabricating 6 samples with $\gamma = 0.15 \text{ cm}^{-1}$ and a coupling of $c = 0.475 \text{ cm}^{-1}$ and a strain that ranges from $1 \leq \tau \leq 2{,}9$. In each sample, we perform 6 single-channel excitations of bulk sites, corresponding to 3 unit cells, and measure the total power remaining in the lattice at the output facet. The extracted data is plotted in Fig. 3. As expected, the variance substantially decreases and tends toward zero as the lattice brought into its unbroken phase

The phase transition from the broken to the unbroken $\mathcal{PT}$-symmetry regime in the graphene lattice is inextricably linked to a topological phase transition related to the emergence of topological mid-gap states. Figure 4a summarizes this feature in a $\tau - \gamma$ phase diagram. The presence or absence of topological mid-gap states at the edge of the Hermitian graphene lattice ($\gamma = 0$) can be reconciled from the perspective of the winding number in a Su-Shrieffer-Heeger (SSH) chain perpendicular to the edges [28, 29]. A topological phase transition from a two-dimensional topological semimetal to a trivial insulator, accompanied by a change of the SSH winding number, takes place at $\tau = 2$ [25,30,31].

For any $\gamma > 0$, however, a topological mid-gap state spontaneously breaks $\mathcal{PT}$-symmetry, since its real dispersive part is pinned. As a consequence, the unbroken $\mathcal{PT}$-symmetric domain of the $\tau - \gamma$ phase diagram does not exhibit any edge modes, as shown in Fig. 4a. In other words, one can either observe unbroken $\mathcal{PT}$-symmetry, or topological mid-gap states, but never both at the same time, since both phenomena exclude each other [32]. Interestingly, a third domain is wedged between the previously discussed cases in the phase diagram. It arises when the strain $\tau$ exceeds the gap threshold determined by $\gamma$. As the gap is closed, $\mathcal{PT}$-symmetry is invariably broken – yet, as long as the hyperboloids in the real part of the dispersion relation still touch, topological mid-gap states are prevented from forming at the bearded edge [33]. This can be intuitively explained by the shape of these states, which reside exclusively within one of the sublattices, and, hence, experience solely $+\gamma$ or $-\gamma$. As a consequence, the imaginary part of the mid-gap dispersion attains $\pm\gamma$ and, from the perspective of a $\mathcal{PT}$-symmetric SSH chain, this implies the disappearance of the mid-gap state for $\tau \geq 2 - \gamma/c$ [12, 33]. Therefore, when starting in the topologically non-trivial domain with mid-gap states and broken $\mathcal{PT}$-symmetry, and following a vertical trajectory for fixed $\gamma$ and increasing $\tau$ in the phase diagram, the system passes not one but two phase transitions. The first occurs when the direct gap of the topological mid-gap states closes at $\tau = 2 - \gamma/c$, which is, hence, of topological nature. The second occurs at $\tau = 2 + \gamma/c$ when the gapped unbroken $\mathcal{PT}$-symmetric domain is reached (see Fig. 4a).

These transitions are exactly what we observe in our experiment. For the unstrained system ($\tau = 1$) and a loss of $\gamma = 0.15$ cm$^{-1}$ such that $\frac{\gamma}{c} = 0.32$, the systems exhibits topological mid-gap states (Fig. 4b left), which we excite by launching light into a waveguide residing at the edge (Fig. 4b right). When increasing the strain $\tau = 2.2$, the mid-gap edge states disappear, in the dispersion relation (Fig. 4c left) as well as in the experiment (Fig. 4c right). However, the system is still in the broken $\mathcal{PT}$-symmetric phase, as shown in Fig. 3b. When the strain is increased to $\tau = 2.9$, the system drives into the unbroken $\mathcal{PT}$-symmetric phase, whereas the mid-gap states are still absent (see Fig. 4d left for the dispersion relation and Fig. 4d right for the experimental data).

In our work, we devised and experimentally demonstrated a two-dimensional $\mathcal{PT}$-symmetric crystal, using an optical platform. To this end, we developed a technology to efficiently introduce artificial losses into the system and unequivocally proved that our realized structure is indeed in the unbroken $\mathcal{PT}$-symmetric phase. Moreover, we highlighted the close connection of a $\mathcal{PT}$-symmetry phase transition to a topological phase transition in our graphene lattice. These findings lay the foundations for realizing two-dimensional $\mathcal{PT}$-symmetry in other wave systems beyond photonics, such as matter waves, sound waves, and possibly even plasmonics and electronic circuits. Moreover, our works opens the gate for future investigations exploring the full potential of $\mathcal{PT}$-symmetric in higher dimensions and may provide the tools to experimentally address numerous exciting questions such as impact of nonlinearity, single photon interference, and many-body effects in two-dimensional $\mathcal{PT}$-symmetric systems.

**Figures:**

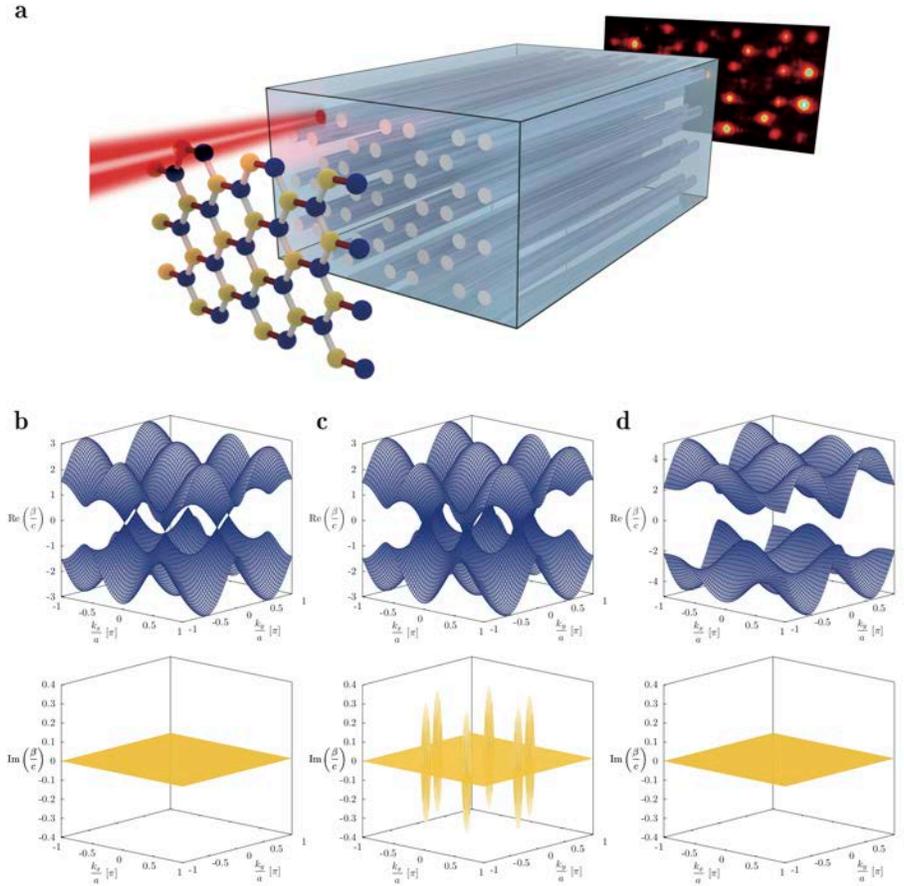

**Fig. 1: The $\mathcal{PT}$-symmetric graphene lattice.** (a) Due to the quantum-optical analogy between solids and waveguide lattices, the probability density of single electron in a carbon graphene lattice shares the same evolution equation as a light beam injected into a honeycomb waveguide lattice. (b) The dispersion relation of a graphene lattice with no gain/loss and no strain ($\tau = 1$), (c) with gain/loss and no strain ($\tau = 1$), and (d) with gain/loss above the critical point ($\tau > 2 + \frac{\gamma}{c}$), which corresponds to the unbroken $\mathcal{PT}$-symmetry regime.

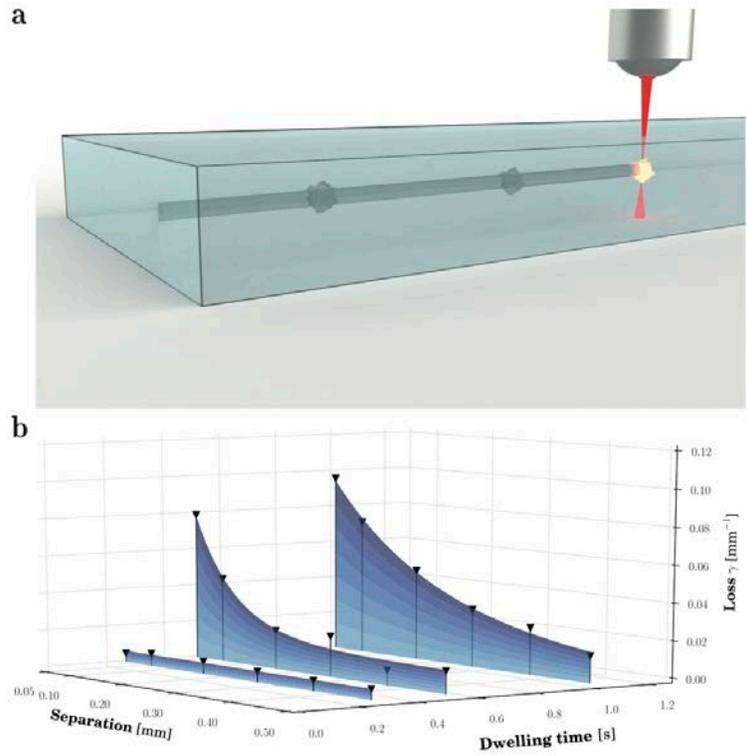

**Fig. 2: Realizing artificial loss.** (a) Schematic of the artificial scatterersintroduced during the fabrication process by dwelling. (b) Experimentally obtainable artificial losses as a function of dwelling time and separation of the scatter centers.

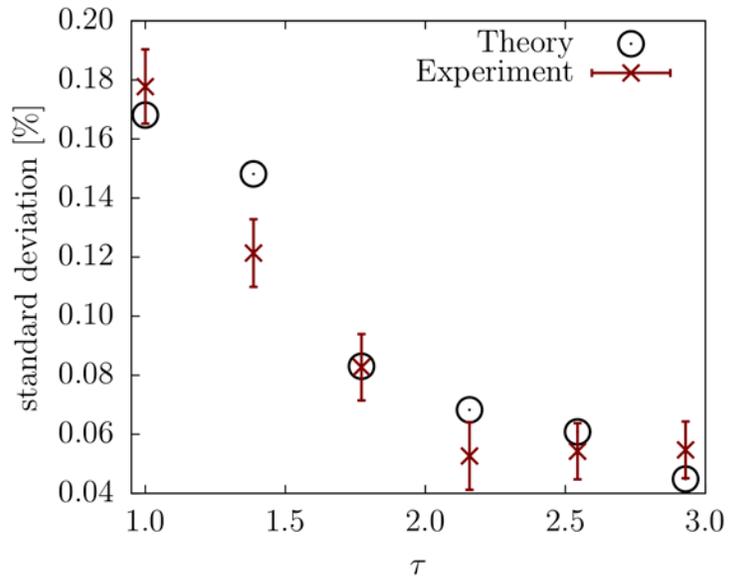

**Fig. 3: Demonstration of the phase transition from broken to unbroken $\mathcal{PT}$-symmetry.** (a) The standard deviation of the output intensity pattern resulting from 6 excitations (3 unit cells). In the unbroken $\mathcal{PT}$-symmetric phase, this quantity approaches zero.

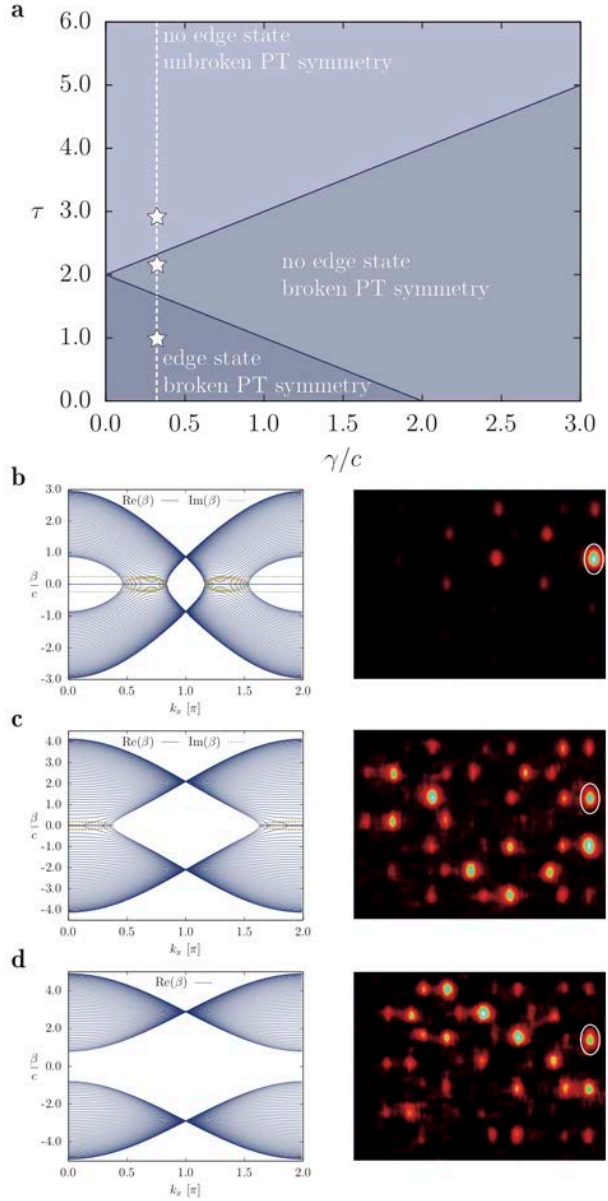

**Fig. 4: Demonstration of the topological transition in the $\mathcal{PT}$-symmetric structure.** (a) The $\tau - \gamma$ phase diagram shows three distinct phases associated with the strain and loss parameters of our structure. (b) For a strain of $\tau < 2 - \frac{\gamma}{c}$, the system is in the broken $\mathcal{PT}$-symmetric phase and topological mid-gap states are present. In the left panel, the corresponding dispersion relation is shown. The right panel presents the experimental image at the sample output facet, where light was injected into the marked edge waveguide and remains close to the edge as the mid-gap state exists. (b) For an intermediate strain $2 - \frac{\gamma}{c} < \tau < 2 + \frac{\gamma}{c}$, the system is still in the broken $\mathcal{PT}$-symmetric phase, but the topological mid-gap states cease to exist. In the left panel, the corresponding

dispersion relation is shown. The right panel presents the experimental image at the sample output facet, where light was injected into the marked edge waveguide and spreads into the bulk as no mid-gap state exists. (b) For sufficiently strong strong strain $\tau > 2 + \frac{\gamma}{c}$, the system is finally driven into the unbroken $\mathcal{PT}$-symmetric phase, where no topological mid-gap states exist either. In the left panel, the corresponding dispersion relation is shown. The right panel presents the experimental image at the sample output facet, where light was injected into the marked edge waveguide and spreads into the bulk as no mid-gap state exists.

## Methods:

The waveguides were manufactured using the femtosecond laser writing method [24] in 10 cm long samples composed of fused silica glass (Corning 7980). The laser pulses are created by a regenerative Ti:Sapphire amplifier system (Coherent RegA seeded with a Mira 900) and exhibit an energy of 450nJ @ 800nm wavelength and 100kHz repetition rate. A highly precise positioning system (Aerotech ALS 130) in concunction with a microscope objective (0.35NA) provides the highly accurate focusing of the laser beam 50 µm to 800 µm under the sample surface. By translating the sample with a speed of about 100mm/min, the refractive index at the focal point is increased by approximately $7\times10^{-4}$, resulting in waveguides with a mode field diameter of 10.4 µm x 8 µm for the 633 nm illumination wavelength. Intrinsic propagation losses and birefringence were estimated to be 0.2dB cm^-1 and $10^{-7}$, respectively. Upon characterization, the output intensity patterns resulting from the excitation of lossy sites were normalized to compensate for the systematically lower injection efficiency.

## Acknowledgements

AS gratefully acknowledge financial support from the Deutsche Forschungsgemeinschaft (grants SZ 276/9-1, SZ 276/19-1, SZ 276/20-1) and the Alfried Krupp von Bohlen und Halbach Foundation. RT is supported by DFG-SFB 1170 (project B04) and ERC-StG-Thomale-TOPOLECTRICS-336012. The Authors would also like to thank C. Otto for preparing the high-quality fused silica samples used in all experiments presented here.

## Author contributions

M.K. and A.S. developed the idea and designed the structure. M.K., R.T. and A.S. worked out the theory. M.K. and T.B. fabricated the samples and performed the measurements. A.S. supervised the project. All authors discussed the results and co-write the paper.

## Additional information

Supplementary information is available in the online version of the paper. Reprints and permissions information is available online at www.nature.com/reprints. Correspondence and requests for materials should be addressed to alexander.szameit@uni-rostock.de.

## Data availability

All experimental data and any related experimental background information not mentioned in the text are available from the authors on request.

**Competing financial interests**

The authors declare no competing financial interests.